\documentclass{ijuc}
\usepackage[pdftex]{graphics}
\newtheorem{theorem}{Theorem}
\usepackage{amsmath}

\begin{document}

\title{A complexity perspective on fluid mechanics}

\author{Saksham Sharma\inst{1,3}\email{ss2531@cam.ac.uk; saksham096@gmail.com}
\and Giulia Marcucci\inst{2} 
\and Adnan Mahmud\inst{3}
}

\institute{Pembroke College, Cambridge, CB2 1RF
\and
Apoha Ltd, London, Acklam Rd, W10 5JJ
\and
Department of Chemical Engineering, University of Cambridge, CB 30 AS, UK
}

\def\received{Received 30th November; In final form }

\maketitle

\begin{abstract}
This article attempts to use the ideas from the field of \textit{complexity sciences} to revisit the classical field of fluid mechanics. For almost a century, the mathematical self-consistency of Navier-Stokes equations has remained elusive to the community of functional analysts, who posed the Navier-Stokes problem as one of the seven millennium problems in the dawn of 21st century. This article attempts to trace back the historical developments of \textit{fluid mechanics} as a discipline and explain the consequences of not rationalising one of the commonly agreed upon tenets - \textit{continuum hypothesis} - in the community. The article argues that `fluids' can be treated as `emergent' in nature, in that the atoms and molecules in the nanometre length scale can likely be correlated with the continuum physics at the microscale. If this is the case, then one might start trying to find a theoretical framework that models the emergence of fluids from atoms, effectively solving the multi-scale problem using a single abstract framework. \textit{Cantor set} with layers $N$ (N can have up to two orders of magnitude) is presented as a potential contender (analytical framework) for connecting the energy in molecular level $C_{1}$ at length scale $l_{cut}$ to the energy at continuum level $C_N$ with length scale $L$. Apart from fluid mechanics, Cantor set is shown to represent the conceptual understanding of VLSI hardware design ($N=5$). Apart from Cantor set, an experimental technique of \textit{architecting metafluids} is also shown to solve emergence experimentally (i.e. connect physics at specific lower scales to higher scales).

\end{abstract}


\section{Introduction}
Matter, as we see it, with naked eyes or with external aids (to name a few, optical lenses, image reconstructions from electromagnetic radiations) can be modeled as existing in three-dimensional space ($\mathcal{R}^{3}$) and amenable to differential calculus treatment. This article focuses on fluid particularly as the matter and reflects on the analytical treatment of its dynamics.\newline 

The primary question of interest is: if one is given a continuum \footnote{} of fluid  in $\mathcal{R}^{3}$, which is external to the observer, does there exist an abstract framework onto which the physics of fluids can be encoded? Is the mathematical framework, built off orthogonal coordinate systems in $\mathcal{R}^{3}$, self-consistent to model the dynamics of fluid? More fundamental question could be: what is ``fluid" fundamentally composed of?  \newline
As philosophical and fundamental these questions might sound, implicit answers to them, commonly accepted in scientific community, have dictated the course of scientific research for past almost two centuries. Starting with Euler, it was conceptualised that the `fluid' is composed of tiny parcels of continuum (of length scale significantly higher - 2 or more orders of magnitude - than molecular length scale) such that the Newton's second laws of motion can be applied in point-wise fashion inside the parcel \cite{euler1757principes, truesdell1960program}. Such a treatment led to the development of `hydraulic engineering', `fluid mechanics', `continuum mechanics' as separate disciplines (that we see now) and the use of calculus in the continua of fluid became more and more accepted \cite{lemarie2018navier}. \newline  

What is seen as modern `fluid mechanics' discipline is equally the result of the ideas which were not accepted by the academic community over generations. A prime example is the `Laplace's demon' which was an optimistic hope of Pierre-Simon Laplace to predict the future of a physical system given its past and the classical laws that if obeys \cite{pierre2007philosophical}. This hope died many deaths with the discovery of laws of thermodynamics, quantum mechanics, and with the construction of Turing machine. To give a sample argument against Laplace's demon, Josef Rukavicka \cite{rukavicka2014rejection} and David Wolpert \cite{wolpert2008physical} noted that the construction of Turing machine and inference devices which perform observation and prediction of the natural world would lead to a fundamental mathematical structure in its logic, called ``halting problem". This problem prevents the existence of a master rule or law that tells when the output of the device would be fixed and determinate, and hence the Laplace's demon is bound to fail. Even after the failure of Laplace's demon, the quest shifted towards connecting the continuum (macroscopic) laws of physics to atomic/molecular contributions. French physicists and mathematicians in the 18th century hugely debated the molecular origins of continuum forces, as discussed by Darrigol \cite{darrigol2002between}. Navier incorporated the molecular forces that are relevant only during the deformation of a fluid, Poisson, on the other hand, summed up the molecular forces by the surrounding molecules acting on a given molecule. Cauchy's formulation was quite similar to Poisson, which is sometimes referred to as the `Cauchy-Poisson' theory. Later on, Poisson offered a fluid theory of motion by treating a fluid like a solid, which experiences stresses during its motion so that the fluid stress is related to the its rate of deformation, which went on to become Navier-Stokes equation with additional pressure gradient term. \newline

The formulation of Navier-Stokes equations as the descriptor of motion of fluids established a boundary between the disciplines: one disciplines is focused on analysing fluids at a continuum level and the other at an atomic/molecular level. However, it does not mean that the two fields are not connected with one another. To give an example, for an ideal gas $G$ at a thermodynamic equilibrium, if $G$ has temperature $T$ and the molecules that combine to form $G$ have average kinetic energy $E$, then $E = k_{B} T N_{DOF} $. Another example, for fluid $F$ with surface tension $\gamma$ composed of molecules in two layers separated at a distance $d_0$, is that the surface energy can be defined as: $\gamma = A / d_0^{6}$ where $d_0$ is the intermolecular distance between two layers.  This means that it is possible, in some cases, to find laws that relation the dynamics at atomic level with the dynamics at continuum level. However, there still is a strong boundary in the level of treatment between continuum mechanics and atomic/molecular physics: in modern era of science, this led to distinction between fluid mechanics and statistical physics as separate disciplines \cite{landau2013fluid, landau2013statistical}. \newline

Though development of distinct disciplines has led to many development in both fronts, it has also led to many unresolved problems in each discipline that has captured attention of physicists for a century or more. For continuum level fluid mechanics, the biggest unresolved problem is the Navier-Stokes regularity problem \cite{fefferman}. In a loose sense, the problem is that of finding a proof that the solutions to Navier-Stokes equations either blow up or remain finite after a certain finite time given a set of initial datum. On the other hand, for statistical physics, one of the unresolved problem is finding an analytical treatment of the dynamics of a large N-particle system, also described by BBGKY set of equations \cite{hamilton1988hierarchical}. It is interesting to note that both problems suffer from a solution \textit{only} because the problem is restricted at the first place to a certain scale of analysis: `continuum' for Navier-Stokes problem and `atomic/molecular' for the BBGKY problem. It would be ideal to have an abstract framework which connects any arbitrary two different physical scales, and bringing more unified picture of the physical problem concerned. \newline

Even without the absence of such a framework, much of the past many decades of research in mathematical fluid dynamics has been influenced by the idea of complexity/emergence. Towards the end of 20th century, Phillip W. Anderson in his much celebrated essay ``More is Different" \cite{anderson1972more} argued the twin difficulties of scale and complexity, when attempting to explain a natural phenomenon in terms of known fundamental laws. In his own words, ``the ability to reduce everything to simple fundamental laws does not imply the ability to start from those laws and reconstruct the universe", meaning that at every new scale a new set of laws emerge. This scientific philosophy of ``complexity science" has pervaded numerous disciplines over the past 50 years, such as origin of life, nonlinear synchronisation, self-assembly in active systems, ants and termites self-organising colonies, and network science, as recently summarised in a recent article \cite{strogatz2022fifty} by eight eminent complexity scientists. On \textit{fluids}, Anderson wrote in his essay that, ``We have already excluded the apparently unsymmetric cases of liquids, gases, and glasses. (In any real sense they are more symmetric.)". The contention here is that in a practical sense, fluids are quite predictable (upon the application of continuum hypothesis) hence, they can be assumed to be symmetric. However, this symmetry holds true as long as the laws of continuum mechanics are attempted to understand at a certain length scale. Any fluid related phenomenon breaks its symmetry and becomes difficult to be modelled when it is seen emergent in nature and one  attempts to find a model that encompasses multiple scales beginning from atomic/molecular contributions. \newline

Apart from emergence, there are few more perspectives posed in the past few decades that challenges our classical treatment of `fluids'. Primary notion that is being challenged is to see mathematical physics (framework to model fluids) as fundamentally continuum in nature; it is not the case in mathematics because of the invent of Turing machines \cite{feferman1988turing} and not in the physics of fluids because the basic units of fluids are discrete set of interacting atoms. There is even more convincing argument that the physics is fundamentally information-theoretic in origin. The primary argument for this is the choice of doing an experiment that yields an observation which eventually lets one record what was not measured by the equipment before. In nutshell, the decision to perform an experiment yields an information about the `fluid' which then yields the governing equations or algorithm describing the physics of the experiment. Such an \textit{observer-participatory} process to yield information and then the physics can offer another way of looking at `fluid mechanics', where instead of treating `fluid' parcels as fundamentally continuum of numbers, one treats them as fundamentally `bits' and discrete chunks of information in nature. If one accepts these arguments, then one is liable to argue `fluid mechanics' as somewhat digital in nature. For past few decades, numerous works are pointing towards the creation of `Digital Fluid Mechanics (DFM)' as a potential discipline which has a potential to thrive amongst existing scientific communities with variety of theoretical, experimental, empirical, and numerical approaches. It should be noted that while machine learning methods can be seen as contenders for empirical and numerical approaches within DFM context, experimental approaches remain the same as conceived before in traditional fluid mechanics research, and theory of neural networks can be considered as the theoretical framework for DFM in this regard \cite{bahri2020statistical}.  \newline

Keeping aside the potential creation of DFM as a discipline and unsolved problems that fluid mechanics and statistical physics are plagued with, there are many experimental observations using fluids that can be discussed as classroom examples of emergence. For example, Couette flow occuring when a fluid is placed between two rotational cylinders at different velocities gives rise to the roll of vortices when the velocity gradient is above a threshold. Another example is the Benard cells formed when a layer of fluid between two horizontal plates is heated from below to lead to the emergence of convection rolls, in the form of multiple attractors - the end result is dependent upon chance fluctuations and is often unpredictable \cite{bishop2008downward}. Over the past one decade, there has been an upsurge in the discovery of nonequilibrium phenomenon centred around the idea of emergence \cite{shankar2022topological}. To give an example, Palacci \textit{et al.} (2013) reported that synthetic photoactivated colloidal particles have a dynamic assembly in the form of periodic crystals that results from the competition between the self-propulsion of particles and attractive interactions caused by the osmotic and phoretic effects \cite{palacci2013living}.  \newline

This main result presented in this article is an abstract framework - a Cantor set - that can be used to connect atomic/molecule scale to continuum scale and present a more unified approach to understanding the dynamics of fluids. The ideas from the field of complexity sciences are introduced in Section 2, along side justifying the need for a framework that can be used to model emergent phenomenon. By drawing focus on fluid mechanics, the unsolved Navier-Stokes regularity problem (discussed in Section 3) is used as a case example which has been attempted in the past by functional analysis communities to be solved using traditional mathematical methods (assuming fluid is composed of continuum at the core) discussed in Section 4. New ideas on potentially solving the Navier-Stokes problem is discussed in Section 6. Section 7 is on Cantor set which can be used to model `fluids' as discrete, information-theoretic and computational in nature composed of atoms/molecules. It is being shown through an example calculation that it is possible to model the emergence of `fluids' from `atoms' by modeling the interaction between atoms composed in different discrete Cantor sets as `rotary logic gates'. Section 8 is on the potential usage of ``Cantor sets" in VLSI hardware design problem (to show its appeal outside the field of physics). Section 9 is on the existing engineering techniques of solving emergent problems without using an abstract framework using physical learning \cite{stern2022learning} and architecting metamaterials \cite{brun2022fluid}. 

\section{Revisiting complexity sciences}

A \textit{system} is a collection of interacting or interdependent components that operate in accordance with a set of rules to produce a holistic entity. Conversely to the conventional definition, a \textit{system} with reference to its individual components, \textit{complexity} or \textit{complex system} defies component-based characterization since no component is independent of the behaviour of the other components. These components establish networks of interactions (often with only a few components involved in numerous interactions) and are capable of generating novel information. Therefore, it is not feasible to strictly formulate the collection's properties based just on knowledge of its components. Hence, the investigation of complex science requires, in general, a new mathematical framework and, in specific, a new sets of measurement metrics. This section of the paper analyses the types of measurement metrics available in the literature for complex systems. \newline
Complex science derives its multidisciplinary nature from the premise that shared properties connect systems across fields and thus justifying the search for universal sets of measuring metrics (modelling techniques in general) applicable to complex systems stemming from all scientific and professional domains covering physics, biology, medicine, engineering, ecology, social sciences, finance, business, management, politics, psychology, anthropology, and more. Due to the multidisciplinary nature of complex science and ubiquitous characteristics such as emergence \footnote{The characteristics of a system that are not evident from its components in isolation, but arise from their interactions, dependencies, or linkages when brought together in a system.} , nonlinearity \footnote{A system in which a change in the input size does not result in a proportional change in the output size.}, and adaptation \footnote{The ability to adapt and gain knowledge from experience.} make defining measuring metrics for complex science discipline-wise is not the optimised method. Moreover, relatively recent work on the analytical framework of complex networks has resulted in a considerable reaffirmation of similarities between complex systems in several scientific fields  \cite{amaral2004complex}. \newline

If the measurement metrics are sectioned from the perspective of the definition of a complex system, then the majority of existing complexity metrics fall into two categories \cite{sporns2007complexity}:

1.	Complexity as Randomness: metrics in this category try to capture the randomness, information content, or description length of a system or process, with random processes being the most complicated since they defy compression the most. This is because compression relies on exploiting a skewed distribution in the data, and randomness and skewness are inversely proportional to each other. \newline

2.	Complexity as Structure and Information: metrics in this category conceptualise complexity as distinct from randomness. In this context, complex systems include a great deal of structure or information, frequently spanning numerous temporal and spatial scales. Within this set of metrics, very complex systems fall halfway between highly ordered (regular) and highly disordered systems (random).\newline

\begin{figure}
\begin{center}
\scalebox{0.5}{\includegraphics{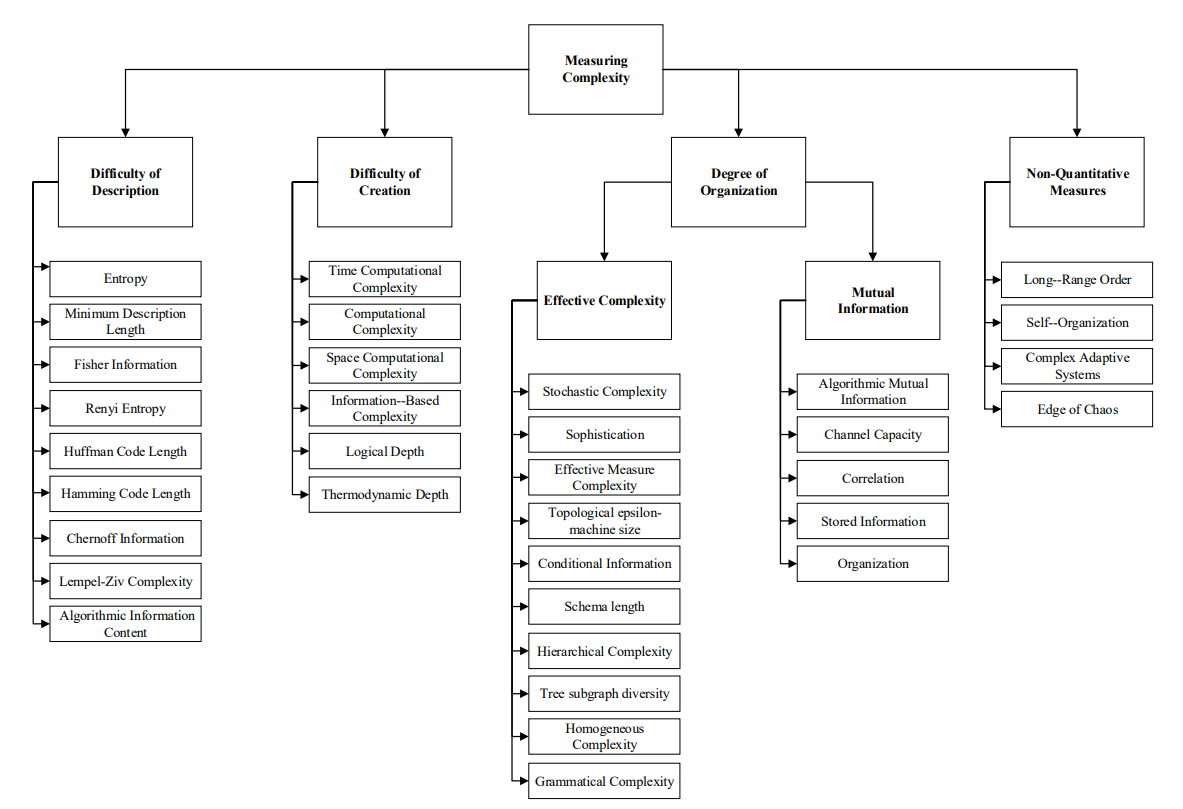}}    
\end{center}
\caption{A non-exhaustive list of measuring metrics for complex system \cite{lloyd2001measures}.}
\label{fig:fig1}
\end{figure}

According to Seth Lloyd \cite{lloyd2001measures}, contemporary researchers were asking the same questions regarding the complexity of their respective research subjects; nonetheless, the solutions they provided for how to assess complexity are quite comparable. Therefore, if the measurement metric is sectioned from the perspective of the questions researchers frequently ask to quantify the complexity, then the metrics can be sectioned into four main categories:
\begin{enumerate}
\item Metrics aiming to measure the \textit{difficulty} in creating a complex system.
\item Metrics aiming to measure the \textit{difficulty} in creating a complex system.
\item Metrics aiming to measure the \textit{degree of organisation} of a complex system. This criterion may further be divided up into two quantities:
 \begin{enumerate}
 \item Effective Complexity: metrics aiming to measure the \textit{difficulty in describing the organisational structure} of a complex system. 
\item Mutual Information: metrics aiming to \textit{evaluate the information shared between the parts} of a complex system as the result of this organizational structure.
\end{enumerate}
\item	Non- quantitative metrics. 

A non-exhaustive list of metrics, sectioned with the latter method is visualised in Figure \ref{fig:fig1}. 

\end{enumerate}

\section{Navier-Stokes regularity problem}
\label{NSR}
Navier-Stokes equations (NSE) are used to model the dynamics of a given fluid given by 
\begin{equation} \label{eq:ns}
\partial_t \vec{u}(t,x) = \nu \Delta \vec{u} - \sum_{i=1}^3 u_i \partial_i \vec{u} -  \vec{\nabla} p + \vec{f} 
\end{equation}
\begin{equation} \label{eq:div}
    \nabla.\vec{u}=\sum^3_{i=1} \partial_i u_i = 0 
\end{equation}
where $\vec{u}(t,x)$ is the velocity vector field in time $t$ and space $x$, $\nu$ is the kinematic viscosity, $\vec{f}$ is the external body force. The initial conditions of the velocity vector field are given as 
\begin{equation} \label{eq:init}
    \vec{u}(0,x)=\vec{u}_0 (x)
\end{equation}
where Eq. \ref{eq:ns} and Eq. \ref{eq:div} are defined for $t \geq 0$ and $x \in \mathcal{R}^n$.

\noindent The problem statement posed by the Clay Mathematical Institute \cite{fefferman} is to prove either that the solutions to Eq. \ref{eq:ns}-\ref{eq:div} remain smooth or they become singular for a given set of initial datum $\vec{u}_{0}(x)$ and after a certain finite-time $T^{*}$.

\section{Efforts to solve the problem}\label{sec:efforts}
Attempts to find solutions to the Navier-Stokes equations (NSE) can be traced back to the first half of the 20th century. In 1911, Oseen \cite{oseen1911formules} used an explicit tensor $\mathcal{O}_{\nu}$, called the \textit{Oseen tensor} to convert NSE into an integro-differential equation  and find the solution
\begin{equation}\label{eq:oseen}
\begin{split}
    \vec{u}(t,x) =& \int_{\mathcal{R}^3}W_{\nu}(t,x-y)\vec{u}_{0}(y) dy + \int_{0}^{t} \int_{\mathcal{R}^3} \mathcal{O}_{nu}(t-s,x-y) \\ &\left(\vec{f}(s,y))-\sum_{i=1}^3u_{i}(s,y)\partial_{i}\vec{u}(s,y)\right) dy  ds
\end{split}
\end{equation}

where $W_{\nu}$ is the heat kernel associated to heat equation $\partial_t T=\nu \Delta T$ where $T$ is temperature. While Oseen derived the solution of \eqref{eq:oseen} for the time interval $[0,T]$, the question of what can be said about possible blowup at $z_{0}$ \footnote{A function $f(z)$ is said to blowup when $f(z) \to \infty$ at $z=z_{0}$. } of such solutions was left unanswered. In 1934, Leray \cite{leray1934mouvement} used an estimate of the energy $\int_{\mathcal{R}^3}|\vec{u}(t,x)|^2 dx$ to find the conditions that prevent the blowup. After realising that the $L^2$ norm \footnote{An $L^2$ norm or the Euclidean norm of a function gives the length of a vector $\mathbf{x}=(x_{1},x_{2},...,x_{n})$ as: $||\mathbf{x}||_{2}=\sqrt{x_{1}^2+x_{2}^2+..+x_{n}^2}$}, for a manifold $\mathcal{R}^n$ is not controllable enough to prevent blowup, Leray introduced a new kind (in fact, concept) of solutions, called \textit{weak solutions}. These solutions have generalised derivatives defined in a Sobolev space $H^1$ \footnote{Sobolev space is the vector space of functions defined by a norm which is the combination of $L^p$ norms and the derivatives of $L^p$ norms to a given order}. Effectively, Leray proved that for an initial datum $u_0 \in L^2$, a global weak solution $\vec{u}(t,x) \in L_{t}^{\infty}L_{x}^2$ exists with an additionally regularity $L_{t}^{2} H_{x}^{1}$. Ever since such a formulation, the fundamental problem of proving uniqueness of the weak solution and that of globalness of strong solutions has remained unsolved. \newline

\noindent Another parameter of paramount importance in NSE analysis is vorticity, $\omega$, given by $\nabla \times u$. In terms of $\omega$, NSE can be written as 
\begin{equation}
    \partial_t \omega + u \cdot \nabla \omega - \nu \Delta \omega = \omega \cdot \nabla u
\end{equation}

where the RHS term is called the \textit{stretching term}. For 2D flow, the RHS is zero because of right angle between the vorticity vector and the flow field plane. As a result, vorticity is controlled in magntiude: if $\nu=0$, vorticity distribution is conserved and $L^p$ norms do not change in time; if $\nu \neq 0$, the norms are non-increasing functions of time. This leads to the regular solutions of NSE and Euler equations, in 2D case, without any formation of singularities. In 3D NSE, the direction of the vorticity vector has an influence on the possibility of the formation of singularities. Constantin,  Fefferman in 1993 \cite{constantin1993direction} proved that if the angle between unit vorticity vectors at positions $x$ and $y$ is $\varphi(x,y,t)$, then for given constants $\Omega$ and $\rho$, if magnitude of the vorticity at both locations is above $\Omega$, then 
\begin{equation}
    \left| \sin \varphi(x,y,t) \right| \leq \frac{|x-y|}{\rho}  
\end{equation}
where the above equality holds true at high values of vorticity and the direction of vorticity is coherent so that the singularities can not form. The physical reason behind this result is the local alignment (vortex tubes being parallel or antiparallel) which regularises the nonlinearity in NSE. \newline

\noindent \eqref{eq:ns} can be collapsed into a seminlinear heat equation
\begin{equation}
    \partial_t u + B(u,u)=\nu \Delta u
\end{equation}
where $B(u,v)$ is the nonlinear quadratic function and a bilinear operator with the general form 
\begin{subequations} \label{semi_heat}
\begin{gather}
    B(u,v) := \frac{1}{2}\mathcal{P}((u.\nabla)v+(v.\nabla)u) \\
    \langle B(u,u),u \rangle = 0
\end{gather}
\end{subequations}
and $\mathcal{P}$ is the orthogonal projection ontp divergence-free vector fields, also called \textit{Leray projection}. Dimensional analysis heuristics can be performed to suggest that NSE might not have much global regularity properties, and are rather supercritical \footnote{Supercritical differential equations have a property that they blowup in finite-time. In hydrodynamics literature, it is used in the sense of supercritical pitchfork or hopf bifurcation.} in nature. Supposing: $\nu=1$; $u(x,t)$ concentrated at a frequency scale, $N(t)$, and amplitude, $A(t)$, gives

\begin{subequations}
\begin{gather}
    \nu \Delta u \approx N(t)^2 A(t) \\
    B(u,u) = \mathcal{P}(u.\nabla u) \approx N(t) A(t)^2 
\end{gather}
\end{subequations}
such that the viscosity dominates nonlinear effects when $A(t) \ll N(t)$ and vice-versa. A power-law model of nonlinear dynamics: $A(t) \approx N(t)^\theta$ for $\theta > 1$ gives 
\begin{subequations}
\begin{gather}
    \partial_t A(t) \approx O(N(t) A(t)^2) \approx O(N(t)^{1+2\theta}) \\
    \partial_t N(t) \approx O(N(t)^{1+\theta})
\end{gather}
\end{subequations}
which alludes that if frequency $N(t)$ is of increasing nature, then the blowup at a finite-time could happen. Now, if the solution $u$ is assumed to exist in a spatial set of intermittency \footnote{In the turbulent fluid flows, intermittency is defined as the pause and start of velocity flow signals. The recorded intermittent signals of the vorticity field $\omega$ is given in \cite[see fig. 2(a)]{lohse_turb}.} dimension $0 \leq \alpha \leq 3$, it should be supported in a ball of volume $N(t)^{-3+\alpha}$. As a result, the energy $\lim_{\mathcal{R}^3} |u(t,x)|^2$ is of the order $\approx A(t)^2 N(t)^{-3+\alpha} \approx N(t)^{-3+\alpha+ 2\theta}$. The energy identity suggests that the exponent of $N(t)$ need to be negative as $N(t) \to \infty $ (blowup), giving the following constraint
\begin{equation}
    \theta \leq \frac{3}{2}-\frac{\alpha}{2}
\end{equation}
alongside the constraint $\theta > 1$ for nonlinear effects to dominate. While high intermittency $\alpha \geq 1$ is not possible for nonlinear effects to dominate \cite{caffarelli1982partial}, if $\alpha < 1$, then blowup is potentially possible. An extreme blowup is possible at $\theta=3/2$ and $\alpha=0$, which means that the solution $u$ is concentrated in a single ball of size $1/N(t)$. Interestingly, there is no result available on the proof that finite time blowup for such a case as $t \to T_{*}<\infty$ is not possible. \newline
A dyadic shell model is often used to model the case: $\theta=3/2, \alpha=0$. Introduced in Katz, Pavlovic \cite{katz2005finite}, the model is given by 
\begin{equation} \label{eq:dyadic}
    \partial_t X_n = \lambda^{n-1}X_{n-1}^2 - \lambda^n X_n X_{n+1}-\lambda^{4n/5}X_n 
\end{equation}

for a constant $\lambda>0$, $X_n$ is the energy of fluid at scale $n-1$ which gets diffused to scale $n$, and the last term in the RHS is the dissipation term. The total energy of the fluid is given by $\sum_n X_n^2$. The analog of blowup in this model is when energy is moved to higher values of $n$, ideally $n \to \infty$. At models corresponding to five and higher dimensions, Cheskidov in 2018 \cite{cheskidov2008blow} showed that when the 4/5 exponent is replaced with an exponent below 2/3, the equations can blowup. However, in 2011, Barbata, Morandin, Romito \cite{barbato2011smooth} proved for a large class of intial datum that the dyadic shell model admitted global regular solutions when $X_n$ exhibited exponential decay as $n \to \infty$, or in other words, are smooth. This was because the energy dispersed too quickly into a broad spectrum of several higher frequency scales, where each frequency mode was activated with amplitude scale such that the effects of viscosity remained stronger in comparison to the nonlinear effects, to prevent blowup and not let the solutions dissipate to infinitely larger frequency scales. \newline

One type of scaling of NSE is given as 
\begin{equation}
    v(x,t) \mapsto v_{\lambda}(x,t)=\lambda v (\lambda x, \lambda^2 t)
\end{equation} 
\noindent and there are various partial regularity results known for this. One of the most prominent one is called Ladyzhenskaya-Prodi-Serrin conditions (\cite{ladyzhenskaya1957existence}, \cite{prodi1959teorema}, \cite{serrin1961interior}), which states that if the Leray weak solution lies in $L^p_t L^q_x$ with $2/p + 3/q \leq 1$, then the solution is unique and smooth in positive time. The endpoint in the above equality for $p=\infty$ and $q=3$, given for $L^{\infty}_t L^3_x$, was proved by  Escauriaza-Seregin-\u Sver\'ak \cite{escauriaza2003img}. It should be noted that while the natural scaling of weak solutions has a regularity for the condition, $2/p + 3/q = 3/2$, the energy equality holds in NSE with an additional regularity for the condition, $2/p + 3/q = 5/4$ \cite{shinbrot1974energy}. This means that there is a gap open between the natural scaling of the equations and the kinetic energy, which can lead to non-uniqueness of weak solutions. In fact, this nonuniqueness was hinted by Jia-\u Sver\'ak \cite{jia2014local}, where the authors proved that in the class $L_t^\infty L_x^{3,\infty}$, Leray solutions are nonunique if a certain spectral assumption holds for a linearised Navier-Stokes operator. In 2019, Buckmaster and Vicol \cite{buckmaster2019nonuniqueness} proved a stronger result that the weak solutions to the 3D NSE, $v \in C_{t}^{0} H_{x}^{\beta}$ (where $H^{\beta}$ is the $L^{2}$ space with the regularity index $\beta$), are nonunique. The nonuniqueness refers to ill-posedness and existence of infinitely many solutions so that a clear statement about the nature of the solution can not be made. The idea used in the proof of this work is the convex integration method \cite{de2015h} and the authors expect that these tools might in the future establish nonuniqueness of Leray weak solutions.
\section{Nature of the problem}
\label{nature}
Interestingly, it is not the first time that such a question - on whether an equation yields physically reasonable answers for an arbitrary set of initial values - has been asked. There have been many problems from other fields in physics and mathematics that resemble to this problem. These include the ray tracing problem in 3D optical systems; recurrent neural networks; nonabelian topological quantum-field theory; stability of $n-$ body systems; and quantum spectral gap problem, ask a similar, \textit{generalised, question}. For example, the quantum gap spectral problem states that given a quantum many-body Hamiltonian, is the system that it describes gapped or gapless? In fact, as recently as 2015, this question was proved to be \textit{undecidable} (in other words, not answerable) \cite{cubitt2022undecidability}. The notion of the undecidability of a system goes back to the times of the Russell paradox - a paradox inherent in the statement ``This statement is false" formulated by Bertrand Russell. He showed that this statement leads to a self-contradiction because the statement predicates itself. Later on, an abstract computing machine, the Turing machine, devised by Alan Turing, was marred by the same self-referential paradox, termed the \textit{Halting problem}. Turing showed that the question of whether for any arbitrary pair of programs and inputs, the machine will run or halt is unanswerable; meaning that no program exists to tell the consistency of all possible pairs of programs and given inputs. Turing replaced the notion of a `question' with a `program' to arrive at the result. It has been more than 80 years since this abstract machine was designed, and even now, the lack of the self-consistency of this machine has had enormous consequences on the self-consistency of various other topics in physics. The quantum gap spectral problem was shown to be unanswerable by formulating it in Turing machine terms, thus bringing in the pathological nature of the machine. In a loose sense, one way to visualise this pathology is by considering an ancient symbol \textit{ouroboros} which is that of a serpent eating its own tail. To ask a question on a problem by asking whether the problem is solvable or not for arbitrary cases is akin to a snake evading or attacking its own tail. This notion of accurate solvability for Navier-Stokes equations (NSE) has been the subject of enquiry for almost a century, long before the Navier-Stokes problem became a Clay problem.

\section{A new scheme to look at the problem}
\label{new_scheme}
In Section \ref{sec:efforts}, numerous efforts to either prove uniqueness or nonuniqueness of the solutions to NSE were outlined. This section, specifically, highlights a new scheme to solve the problem by using ideas from theoretical computer science. \newline
\noindent In 2015, Terence Tao \cite{tao2016finite} introduced an improved variant to the dyadic shell model by shunning off most of the nonlinear terms and incorporating only one pair of adjacent modes $X_{n}, X_{n+1}$ which experience a nonlinear and energy interactions at an instant. An improved version of \ref{eq:dyadic} is given as
\begin{equation}
\begin{gathered} \label{eq:dyadic_var}
\partial_t X_n = - \lambda^{2n\alpha} X_n + 1_{n-1=n(t)}\lambda^{n-1}X^{2}_{n-1} \\ - 1_{n=n(t)}\lambda^{n=n(t)}\lambda^{n}X_{n}X_{n+1}
\end{gathered}
\end{equation}
where $n: [0,T_{*})\to\mathcal{Z}$ is a piecewise constant function that describes the mode pair ($X_{n(t)},X_{n(t)+1}$) allowed to interact at a given time $t$. A system of ODEs were constructed to \textit{simulate} \ref{eq:dyadic_var} by using a sequence of ``quadratic circuits" connected in series fashion. Each circuit is composed of ``quadratic logic gates" which simulate a certain form of basic quadratic non-linear interaction. As shown in Fig. \ref{fig:fig3}(c), the gates transfer energy from one mode to another with ``amplifier" and ``rotor" gates. Combination of these gates with certain coupling constants aid in energy transfer from frequency scale $n$ to $n+1$. A bootstrap argument is used to construct a blowup solution to the truncated system of ODEs, interpreted as ``averaged Navier-Stokes equations", concisely given by
\begin{equation} \label{eq:avg_ns}
    \partial_t u + \Tilde{B}(u,u)=\nu \Delta u
\end{equation}
and exhibit blowup for values close to $\theta_{3/2}$ and $\alpha=0$ scenarios discussed previously. It is also important to note that the the dynamics of \ref{eq:avg_ns} is approximately and discretely self-similar in time. Also, noteworthy is that the perfectly continuous self-similar solutions to NSE are not possible, as proved in Necas-Ruzicka-Svreak \cite{nevcas1996leray} and Escuriaza-Seregin-Sverak \cite{escauriaza2003img}. When the viscosity effects are of the lower order, such as in the case of $\theta=3/2$ and $\alpha=0$, viscosity can been a perturbative term. As a result, instead of the NSE, the spotlight then is on the Euler equations ($\nu=0$), specifically ones when the blowup solutions were found. When one considers this equation, Kelvin's circulation theorem is given by 
\begin{equation}
    \Gamma(t)= \int_{C} \textbf{u}.dn =\int_{S} \omega 
\end{equation}
which means that the line integral of $\vec{u}$ around a curve $C$ that loops the fluid is equivalent to the surface integral of the curl of velocity field (vorticity). Euler equations written in terms of vorticity is given as 
\begin{subequations}
\begin{gather}
    \partial_t \omega + \mathcal{L}_u \omega = 0 \\
    u = T \omega
\end{gather}
\end{subequations}
where $\mathcal{L}$ is the Lie derivative \footnote{Lie derivative represents the change in the tensor field, given by the directional derivatives of each component, along the flow fields which are defined by another vector field.} w.r.t $u$ and $T \omega = -\nabla \times \Delta^{-1}(* \omega)$ is the Biot-Savart operator \footnote{In fluid mechanics, for the equations: $\nabla \times u = \omega$, $\nabla.u=0$, the Biot-Savart operator $T$ maps $\omega$ into $\vec{u}$.} This formulation is a concise encoding of the conservation laws that Euler equations exhibit. Tao in 2016 \cite{tao2016finite} constructed a blowup for artificial version of Euler equations where the Bio-Savart operator is replaced by a truncated linear operator of the form $\Tilde{T}$. The blowup scenario  is called a ``vortex neck pinch'' where the streamlines of the vortices are pinched into a ring of radius $O((T_{*}-T)^{1/2})$ (Fig. 2) when the blowup is approached at time $T_{*}$, with $u \sim (T_{*}-t)^{-1/2}$ and $\omega \sim (T_{*}-t)^{-1}$. The blowup is similar to the discretely self-similar NSE blowup scenario with $\theta=1, \alpha=0$. The pinchoff shown in this artificial construction of Euler equations is similar to the those observed in numerical studies \cite{mckeown2020turbulence}.\newline

\newtheorem{conj}{Conjecture}
Given that there is some possibility that the Euler equations can exhibit blowup in finite blowup, a conjecture was postulated by Tao stating that 
\begin{conj}\label{conj1}
There exists some smooth manifold $M=(M,g)$ (of unspecified dimension) and a smooth solution $\vec{u}(x,t),p$ to the Euler equations (with suitable decay at infinity) that blows up in finite time.
\end{conj}
The potential route to prove \eqref{conj1} is by carefully designing the metric $g$ and the dimension $M$ so as to ``program'' various types of behavior that enables blowup. It should be carefully noted that no corners or boundaries should exist in the domain, that might yield a blowup, and ideally, fluid in an infinite and smooth domain. The rational behind designing the manifold is akin to constructing a computer program that operates on the fluid and cause it to blowup. \newline

There are two important features that are required in the solutions to the NSE. The first one is \textit{universality} and the second one is the \textit{scalar invariance}. In 2017, Tao proved the following theorem \cite{tao2017universality}
\begin{theorem}
Let $\partial_t u = B(u,u)$ be an ODE such that $B: V \times V \to V$ is a bilinear form and $V$ is a finite-dimensional inner product space. The conservation law is: $\langle B(u,u),u = 0 \rangle$. Then there exists a compact Riemannian manifold $(M,g)$ and a linear isometry $T: V \to C^{\infty}(M)$ that maps solutions to $\partial_t u = B(u,u)$ to solutions to the Euler equation on M.
\end{theorem}
where a symmetry reduction from finite-dimensional ODE to Euler equations is performed. For a specific $V$, there exists a manifold such that $V$ is embeddable into smooth vector fields. Once this is possible, then any ODE of the above type can be translated to a smooth vector field and then the manifold can be created such that  it exhibits finite time blowup, albeit for fixed number of frequency scales. In 2019, Tao proved another theorem \cite{tao2019universality} which establishes further than Euler equations (or flows) can be seen as universal. The theorem states that 
\begin{theorem}
If $d \geq 2$, then there is a somewhere dense set (in the smooth topology) of incompressible flows $X:[0,T] \to \Gamma (T(R/Z)^{d})$ which can be lifted to an Eulerisable flow on a warped product of $(R/Z)^{d}$ and another torus.
\end{theorem}
where vector field is \textit{Eulersiable} if $X: [0,T] \to \Gamma (T(R/Z)^{d})$, meaning that it solves the Euler equations on manifold $M$ with some metric $g$. The theorem, in summary, proved that in a torus, all vector fields, though not Eulerisable, some of them can be lifted up and mapped to a warped product \footnote{Warped product is a Riemannian manifold whose geometry can be decomposed into a cartesian product of the $y$ and $x$ geometry where the term containing $x$ is warped and is rescaled by a function of the other coordinate $y$.} of the torus such that the flows become Eulerisable. The outcome of these two theorems is the first of many works by Cardona, Miranda, Peralta-Salas, Presas \cite{cardona2019universality} where the authors proved that 
\begin{theorem}
Every geodesible vector field $X$ on a compact manifold $M$ can be obtained as the pullback $X=\phi^{*}Y$ of a stationary Eulerisable flow $Y$ on another manifold $N$ with respect to some embedding $\phi: M \to N$. 
\end{theorem}
where the stationary vector fields (geodesic fields \footnote{Geodesic fields on a Riemannian manifold are the vector fields having integral curves that are geodesics, meaning, the curves that connect two points on a surface by the shortest path.}) can be mapped to stationary solutions to the Euler equations, by embedding the manifold $M$ to bigger manifold $N$. Such flows are also called \textit{Beltrami flows} and they are related to Reeb vector fields using ideas from contact geometry. In 2021, Torres de Lizaur \cite{lizaur2022chaos} proved that 
\begin{theorem}
Given any vector field $X$ on a compact manifold $N$, there exists a Riemannian manifold (M,g) and an invariant manifold $\Tilde{N}$ diffeomorphic to $N$ in the phase space $\Gamma(TM)$ of the Euler equations on $(M,g)$, such that the Euler flow on $\Tilde{N}$ is arbitrarily close to $X$ in the smooth topology after identifying $\Tilde{N}$ with $N$.
\end{theorem}
such that the perturbed vector field in a manifold $\tilde{N}$ can be embdedded in a higher dimensional manifold $N$ such that the solutions to Euler equations on $M$ are quite close to those in $N$ which is stable upon perturbation. One corollary of this work is that there are manifolds $(M,g)$ such that Euler dynamics in these manifolds are chaotic in that they exhibit horseshoe map, homoclinic orbits, and similar features of a chaotic system which are stable upon perturbations. In 2021, Cardona, Miranda, Peralta-Salas, Presas \cite{cardona2021constructing} proved that 
\begin{theorem}
There exists a Riemannian manifold $(M,g)$ whose Euler flow is Turing-complete: the halting problem for any given Turing machine is equivalent to the Euler flow for a certain initial condition associated to that machine entering a certain region.
\end{theorem}
where the authors constructed solutions for the steady Euler flow (without viscosity) on a Riemannian 3-sphere $\mathcal{S}^{3}$ that are Turing-complete. Here, Turing completeness means that for given points on the sphere, the problem of bounding the dynamical trajectory of those points, as the equation evolves, is an undecidable problem. Following this work, the authors extended the analysis to a standard Euclidean 3-dimensional space. The tools used to prove such a result are contact topology, symplectic geometry, h-principle, generalised shift maps, and Cantor sets. 
\begin{figure}
\begin{center}
\scalebox{0.8}{\includegraphics{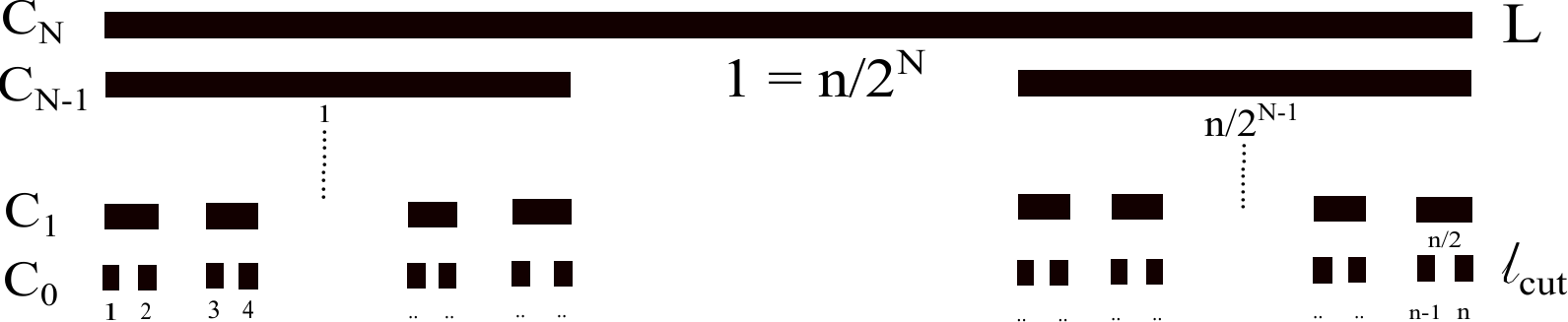}}    
\end{center}
\caption{Cantor set framework consisting of $N$ layers with length scale $L$ of the top-most one and $l_{cut}$ of the bottom-most one. There are $n$ discrete elements in the bottom-most layer, and the number of elements reduce by a factor of 2 as one goes bottom-up layer by layer. Each layer is represented by $C_i$, where $0\leq i \leq N$.}
\label{fig:fig2}
\end{figure}

\section{Cantor set and rotor gate analysis}
We are interested to analyse two length scales: $l_{cut}$ and $L$, such that $l_{cut}<<L$ (lower by few orders of magnitude). For example, for a liquid drop of radius 1 mm sitting on a hydrophobic surface, $L=1$ mm and $l_{cut}=10^{-9}$ nm; $l_{cut}$ is at a molecular scale, $L$ is at a continuum scale. Consider that all the atoms at the length scale $l_{cut}$ can be categorised and put into a collection of countably infinite sets - called Cantor sets - denoted by $\mathcal{C}_{k} \, (1 \leq k \leq N)$. $N$ is the number of layers between $l_{cut}$ and $L$, so that at the length scale $L$, all the atoms can be collected and put in \textit{single} Cantor set denoted by $\mathcal{C}_{N}$. From Fig. \ref{fig:fig1}, it is apparent, geometrically, that the sets at layer $k$ merge pairwise, to form the sets at layer $k+1$. Thus, if the number of subsets in $C_{1}$ are $n$, the the number of subsets in $C_{2}$ are $n/2$, and so on, until $C_{N}$. Since $C_{N}$ is a single set with just one subset (itself), $n/2^{N}=1$. Rearranging it
\begin{equation}\label{eq:law}
    N = \log n_{2}
\end{equation}
At a general length scale, say $l$ ($l_{cut}\leq l \leq L$), with the atoms categorised in the set $C_{k}$, let us call each subset in $C_{k}$ as a ``mode". Consider any two adjacent modes in $C_{k}$, say $x$ and $y$ with energies $e_{x}$ and $e_{y}$. At an initial time step $t_{0}$, the modes are represented by $x(t_{0})$ and $y(t_{0})$. Consider them rotating around the origin at a constant angular rate $\alpha z(t_{0})$ such that 
\begin{equation}
    x(t)=x(t_{0}) \cos(\alpha z(t_{0})) (t-t_{0}) - y(t_{0}) \sin (\alpha z(t_{0})(t-t_{0}))
\end{equation}
\begin{equation}
    y(t)=y(t_{0}) \cos(\alpha z(t_{0})) (t-t_{0}) + x(t_{0}) \sin (\alpha z(t_{0})(t-t_{0}))
\end{equation}
\begin{equation}
    z(t)=z(t_{0})
\end{equation}
where the mode $z \in C_{k+1}$ is fixed while $x,y \in C_{k}$ is not. In fact, starting at $t_{0}$ until time $T_{k}$, the two modes $x$ and $y$ interchange energy with each other and thus mix to $emerge$ and form the mode $z$ at the layer $C_{k+1}$. This process can be viewed as $z$ mode driving the exchange of energy between $x$ and $y$ modes. It is to be noted that though it is the choice of the reader to see the Cantor set in either top-down or bottom-up fashion, throughout this article, we choose to discuss \textit{emergence} as a \textit{bottom-up} process, starting from $l_{cut}$ and reaching layer-by-layer until $L$. \newline
\noindent Equipartition theorem in statistical mechanics gives us
\begin{equation}
    \alpha z (x^{2}-y^{2})=\partial_{t}(xy)
\end{equation}
which can be written as 
\begin{equation}
    \alpha \int^{T_{k}}_{t_{0}} z(t)(x^{2}-y^{2})dt = x(T_{k})y(T_{k})-x(t_{0})y(t_{0})
\end{equation}
such that $x^{2}+y^{2}=E$ and constant $z$ gives 
\begin{equation}
    \frac{1}{T-T_{0}}\int^{T_{k}}_{t_{0}}x^{2} (t) dt = \frac{E}{2} + O(\frac{E}{\alpha |z(t_{0}|)(T-T_{0})|})
\end{equation}
where $E=e_{x}+e_{y}$ and $e_{x}=e_{y}$. As a result, 
\begin{equation}\label{eq:energy_form}
    e_{x}=e_{y} = e[C_{k}] \approx \frac{e[C_{k+1}]}{2}
\end{equation}
Using Eq.\ref{eq:energy_form} and Eq.\ref{eq:law}, one gets the general formula
\begin{equation}
    e[C_{N}] \approx 2^{N} e[C_{1}]
\end{equation}
which relates the energy of a discrete block in $C_{1}$ (molecular scale) to the energy in $C_{N}$ (continuum scale).

\begin{figure}
\begin{center}
\scalebox{2}{\includegraphics{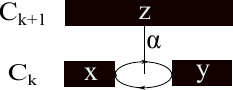}}
\end{center}
\caption{A rotor gate present between 2 layers $C_k$ and $C_{k+1}$ such that $x$ and $y$ modes operate to start dynamics in the $z$ mode, which we term as \textit{emergence} of physics at layer $k+1$ from layer $k$.}
\label{fig:fig3}
\end{figure}
\section{Different abstraction levels in VLSI hardware design as cantor set}
Cantor set as a framework can be used to model and understand phenomenon exhibiting emergence in disciplines other than physics. This section is on the VLSI hardware design problem and how it can be modelled using Cantor set framework. \newline

In 1964, an American semiconductor company, General Microelectronics (GMe) manufactured first-ever MOS (metal-oxide-semi conductor) integrated chips which composed of more than 10,000 transistors in a single chip. Later on, VLSI (very large scale integrated) chips with millions and billions of MOS transistors on them became the core element of semi-conductor industry.\newline

\begin{figure}
\begin{center}
\scalebox{0.7}{\includegraphics{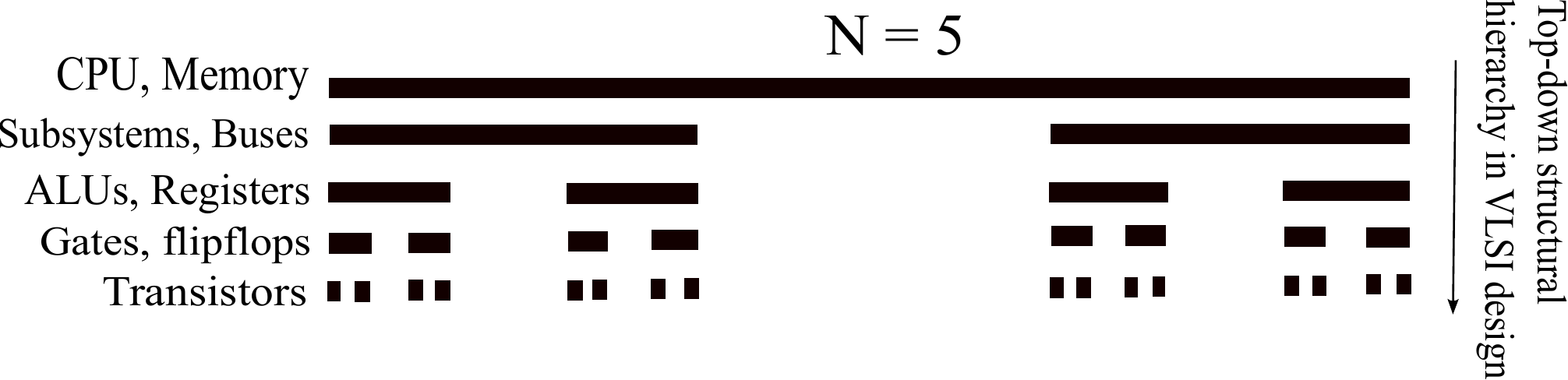}}
\end{center}
\caption{VLSI hardware design top-down hierarchy represented in the form of a Cantor set. There are 5 layers in the structural hierarchy and each of the layer operates in a fashion such that the knowledge of individual element at a certain layer \textit{can not} help to deduce the knowledge of an arbitrary element at the layer above. }
\label{fig:fig4}
\end{figure}

One of the big challenges in the industry is something called, top-design problem, which is to efficiently divide the task of making VLSI chips in different categories. Gajski-Kuhn chart \cite{gajski1983new} is one visual description that identifies three different domains of behaviour - behaviour, structure, layout - that  goes top-down to more refined abstraction levels. It is apparent from the Fig. \ref{fig:fig4} that these domains, structural, can be expressed \textit{conceptually} in a Cantor set framework with N = 5. The interpretation from this representation is that the knowledge and the workforce focused on the transistor level \textit{can not} individually contribute anything of value to the knowledge and workforce at one level above (gates, flipflops). Same argument can be applied for the layers above. It is the collective knowledge at one level that is of use to a level above. This is what is emergence, the \textit{impossibility} of using information from single unit to say anything meaningful of the layer above. However, it is to noted that Cantor set framework for VLSI hardware design is only a conceptual representation, not a quantitative one. 

\section{If not cantor set framework, then what?} 
While the notion of correlating arbitrary lower scales to higher scale might seem to be theoretical in nature, it can also be looked from an engineering and design lens. Recently, Song \textit{et al.} (2022) \cite{song2022microscopic} showed that the linear viscoelasticity of arrested soft materials (gels, glasses) at a macroscopic level is correlated to the microscopic dynamics of the material. At the macroscopic level, the stress relaxation of the system is studied using rheometry techniques. On the other hand, at the microscopic level, the superdiffusive dynamics of microscopic clusters are studied by perturbing the material using X-ray photo correlation spectroscopy. The nonlinear phenomenon at the smaller scale is physically correlated to the linear phenomenon at the macroscopic scale. An experimental understanding of this correlation can inform design experts to interfere and perturb the lower scale phenomenon in such a way that the desired output at the macroscopic level is reached. The techniques of \textit{physical learning} (supervised or unsupervised) \cite{stern2020supervised,stern2020continual} could be useful here to train and adapt a physical system to reach a desired target property by experimentally perturbing its internal nodes and adjusting the weights of the network in accordance with the physical law governing the emergent process. 
\section*{Dedication}
This article is dedicated to Sir George Gabriel Stokes, a physicist who made seminar contributions in the field of fluid mechanics. He served as the Lucasian Professor of Mathematics from 1849 until 1903, and served as the Master of Pembroke College from 1902-1903, at the University of Cambridge. The article written above is the result of never-ending source of inspiration that S.S. received by glancing fondly over the magnanimous portrait of Stokes in the dining hall of Pembroke.
\bibliographystyle{apalike}
\bibliography{ijuc}

\begin{thebibliography}{10}

\bibitem{amaral2004complex}
Luis~AN Amaral and Julio~M Ottino.
\newblock (2004).
\newblock Complex networks.
\newblock {\em The European physical journal B}, 38(2):147--162.

\bibitem{anderson1972more}
Philip~W Anderson.
\newblock (1972).
\newblock More is different: broken symmetry and the nature of the hierarchical
  structure of science.
\newblock {\em Science}, 177(4047):393--396.

\bibitem{bahri2020statistical}
Yasaman Bahri, Jonathan Kadmon, Jeffrey Pennington, Sam~S Schoenholz, Jascha
  Sohl-Dickstein, and Surya Ganguli.
\newblock (2020).
\newblock Statistical mechanics of deep learning.
\newblock {\em Annual Review of Condensed Matter Physics}, 11(1).

\bibitem{barbato2011smooth}
David Barbato, Francesco Morandin, and Marco Romito.
\newblock (2011).
\newblock Smooth solutions for the dyadic model.
\newblock {\em Nonlinearity}, 24(11):3083.

\bibitem{bishop2008downward}
Robert~C Bishop.
\newblock (2008).
\newblock Downward causation in fluid convection.
\newblock {\em Synthese}, 160(2):229--248.

\bibitem{brun2022fluid}
Pierre-Thomas Brun.
\newblock (2022).
\newblock Fluid-mediated fabrication of complex assemblies.
\newblock {\em JACS Au}.

\bibitem{buckmaster2019nonuniqueness}
Tristan Buckmaster and Vlad Vicol.
\newblock (2019).
\newblock Nonuniqueness of weak solutions to the navier-stokes equation.
\newblock {\em Annals of Mathematics}, 189(1):101--144.

\bibitem{caffarelli1982partial}
Luis Caffarelli, Robert Kohn, and Louis Nirenberg.
\newblock (1982).
\newblock Partial regularity of suitable weak solutions of the navier-stokes
  equations.
\newblock {\em Communications on pure and applied mathematics}, 35(6):771--831.

\bibitem{cardona2019universality}
Robert Cardona, Eva Miranda, Daniel Peralta-Salas, and Francisco Presas.
\newblock (2019).
\newblock Universality of euler flows and flexibility of reeb embeddings.
\newblock {\em arXiv preprint arXiv:1911.01963}.

\bibitem{cardona2021constructing}
Robert Cardona, Eva Miranda, Daniel Peralta-Salas, and Francisco Presas.
\newblock (2021).
\newblock Constructing turing complete euler flows in dimension 3.
\newblock {\em Proceedings of the National Academy of Sciences},
  118(19):e2026818118.

\bibitem{cheskidov2008blow}
Alexey Cheskidov.
\newblock (2008).
\newblock Blow-up in finite time for the dyadic model of the navier-stokes
  equations.
\newblock {\em Transactions of the American Mathematical Society},
  360(10):5101--5120.

\bibitem{constantin1993direction}
Peter Constantin and Charles Fefferman.
\newblock (1993).
\newblock Direction of vorticity and the problem of global regularity for the
  navier-stokes equations.
\newblock {\em Indiana University Mathematics Journal}, 42(3):775--789.

\bibitem{cubitt2022undecidability}
Toby Cubitt, David Perez-Garcia, and Michael~M Wolf.
\newblock (2022).
\newblock Undecidability of the spectral gap.
\newblock In {\em Forum of Mathematics, Pi}, volume~10. Cambridge University
  Press.

\bibitem{darrigol2002between}
Olivier Darrigol.
\newblock (2002).
\newblock Between hydrodynamics and elasticity theory: the first five births of
  the navier-stokes equation.
\newblock {\em Archive for History of Exact Sciences}, 56(2):95--150.

\bibitem{de2015h}
Camillo De~Lellis and L{\'a}szl{\'o} Sz{\'e}kelyhidi.
\newblock (2015).
\newblock On h-principle and onsager’s conjecture.
\newblock {\em Eur. Math. Soc. Newsl}, 95:19--24.

\bibitem{escauriaza2003img}
Luis Escauriaza, Gregory~A Seregin, and Vladimir Sverak.
\newblock (2003).
\newblock L3 infinity solutions of the navier-stokes equations and backward
  uniqueness.
\newblock {\em Russian Mathematical Surveys}, 58(2):211--250.

\bibitem{euler1757principes}
Leonhard Euler.
\newblock (1757).
\newblock Principes g{\'e}n{\'e}raux du mouvement des fluides.
\newblock {\em M{\'e}moires de l'acad{\'e}mie des sciences de Berlin}, pages
  274--315.

\bibitem{feferman1988turing}
Solomon Feferman.
\newblock (1988).
\newblock Turing in the land of o (z).
\newblock In {\em A half-century survey on The Universal Turing Machine}, pages
  113--147.

\bibitem{fefferman}
Charles~L Fefferman.
\newblock (2000).
\newblock Existence and smoothness of the navier-stokes equation.
\newblock {\em The millennium prize problems}, 57:67.

\bibitem{gajski1983new}
Daniel~D Gajski and Robert~H Kuhn.
\newblock (1983).
\newblock New vlsi tools.
\newblock {\em Computer}, 16(12):11--14.

\bibitem{hamilton1988hierarchical}
AJS Hamilton.
\newblock (1988).
\newblock On hierarchical solutions to the bbgky hierarchy.
\newblock {\em The Astrophysical Journal}, 332:67--74.

\bibitem{jia2014local}
Hao Jia and Vladim{\'\i}r {\v{S}}ver{\'a}k.
\newblock (2014).
\newblock Local-in-space estimates near initial time for weak solutions of the
  navier-stokes equations and forward self-similar solutions.
\newblock {\em Inventiones mathematicae}, 196(1):233--265.

\bibitem{katz2005finite}
Nets Katz and Nata{\v{s}}a Pavlovi{\'c}.
\newblock (2005).
\newblock Finite time blow-up for a dyadic model of the euler equations.
\newblock {\em Transactions of the American Mathematical Society},
  357(2):695--708.

\bibitem{ladyzhenskaya1957existence}
OA~Ladyzhenskaya and AA~Kiselev.
\newblock (1957).
\newblock On the existence and uniqueness of the solution of the nonstationary
  problem for a viscous incompressible fluid.
\newblock {\em Izv. Akad. Nauk SSSR Ser. Mat}, 21(19579):665--680.

\bibitem{landau2013fluid}
Lev~Davidovich Landau and Evgenii~Mikhailovich Lifshitz.
\newblock (2013).
\newblock {\em Fluid Mechanics: Landau and Lifshitz: Course of Theoretical
  Physics, Volume 6}, volume~6.
\newblock Elsevier.

\bibitem{landau2013statistical}
Lev~Davidovich Landau and Evgenii~Mikhailovich Lifshitz.
\newblock (2013).
\newblock {\em Statistical Physics: Volume 5}, volume~5.
\newblock Elsevier.

\bibitem{lemarie2018navier}
Pierre~Gilles Lemari{\'e}-Rieusset.
\newblock (2018).
\newblock {\em The Navier~ Stokes Problem in the 21st Century}.
\newblock Chapman and Hall/CRC.

\bibitem{leray1934mouvement}
Jean Leray.
\newblock (1934).
\newblock Sur le mouvement d'un liquide visqueux emplissant l'espace.
\newblock {\em Acta mathematica}, 63(1):193--248.

\bibitem{lizaur2022chaos}
Francisco Torres~de Lizaur.
\newblock (2022).
\newblock Chaos in the incompressible euler equation on manifolds of high
  dimension.
\newblock {\em Inventiones mathematicae}, 228(2):687--715.

\bibitem{lloyd2001measures}
Seth Lloyd.
\newblock (2001).
\newblock Measures of complexity: a nonexhaustive list.
\newblock {\em IEEE Control Systems Magazine}, 21(4):7--8.

\bibitem{lohse_turb}
Detlef Lohse and Siegfried Grossmann.
\newblock (1993).
\newblock Intermittency in turbulence.
\newblock {\em Physica A: Statistical Mechanics and its Applications},
  194(1-4):519--531.

\bibitem{mckeown2020turbulence}
Ryan McKeown, Rodolfo Ostilla-M{\'o}nico, Alain Pumir, Michael~P Brenner, and
  Shmuel~M Rubinstein.
\newblock (2020).
\newblock Turbulence generation through an iterative cascade of the elliptical
  instability.
\newblock {\em Science advances}, 6(9):eaaz2717.

\bibitem{nevcas1996leray}
Jindvrich Nevcas, Michael Rvzicka, and Vladimir Sverak.
\newblock (1996).
\newblock On leray's self-similar solutions of the navier-stokes equations.
\newblock {\em Acta Mathematica}, 176(2):283--294.

\bibitem{oseen1911formules}
CW~Oseen.
\newblock (1911).
\newblock Sur les formules de green g{\'e}n{\'e}ralis{\'e}es qui se
  pr{\'e}-sentent dans l'hydrodynamique et sur quelques.
\newblock {\em Acta mathematica}, 34:205.

\bibitem{palacci2013living}
Jeremie Palacci, Stefano Sacanna, Asher~Preska Steinberg, David~J Pine, and
  Paul~M Chaikin.
\newblock (2013).
\newblock Living crystals of light-activated colloidal surfers.
\newblock {\em Science}, 339(6122):936--940.

\bibitem{pierre2007philosophical}
Marquis De~Laplace Pierre-Simon.
\newblock (2007).
\newblock {\em A Philosophical Essay on Probabilities}.
\newblock Cosimo, Inc.

\bibitem{prodi1959teorema}
Giovanni Prodi.
\newblock (1959).
\newblock Un teorema di unicita per le equazioni di navier-stokes.
\newblock {\em Annali di Matematica pura ed applicata}, 48(1):173--182.

\bibitem{rukavicka2014rejection}
Josef Rukavicka.
\newblock (2014).
\newblock Rejection of laplace's demon.
\newblock {\em The American Mathematical Monthly}, 121(6):498--498.

\bibitem{serrin1961interior}
James Serrin.
\newblock (1961).
\newblock {\em On the interior regularity of weak solutions of the
  Navier-Stokes equations}.
\newblock Mathematics Division, Air Force Office of Scientific Research.

\bibitem{shankar2022topological}
Suraj Shankar, Anton Souslov, Mark~J Bowick, M~Cristina Marchetti, and Vincenzo
  Vitelli.
\newblock (2022).
\newblock Topological active matter.
\newblock {\em Nature Reviews Physics}, 4(6):380--398.

\bibitem{shinbrot1974energy}
Marvin Shinbrot.
\newblock (1974).
\newblock The energy equation for the navier--stokes system.
\newblock {\em SIAM Journal on Mathematical Analysis}, 5(6):948--954.

\bibitem{song2022microscopic}
Jake Song, Qingteng Zhang, Felipe de~Quesada, Mehedi~H Rizvi, Joseph~B Tracy,
  Jan Ilavsky, Suresh Narayanan, Emanuela Del~Gado, Robert~L Leheny, Niels
  Holten-Andersen, {\it et~al.}
\newblock (2022).
\newblock Microscopic dynamics underlying the stress relaxation of arrested
  soft materials.
\newblock {\em Proceedings of the National Academy of Sciences},
  119(30):e2201566119.

\bibitem{sporns2007complexity}
O~Sporns, (2007).
\newblock Complexity. scholarpedia, 2 (10), 1623.

\bibitem{stern2020supervised}
Menachem Stern, Chukwunonso Arinze, Leron Perez, Stephanie~E Palmer, and Arvind
  Murugan.
\newblock (2020).
\newblock Supervised learning through physical changes in a mechanical system.
\newblock {\em Proceedings of the National Academy of Sciences},
  117(26):14843--14850.

\bibitem{stern2022learning}
Menachem Stern and Arvind Murugan.
\newblock (2022).
\newblock Learning without neurons in physical systems.
\newblock {\em arXiv preprint arXiv:2206.05831}.

\bibitem{stern2020continual}
Menachem Stern, Matthew~B Pinson, and Arvind Murugan.
\newblock (2020).
\newblock Continual learning of multiple memories in mechanical networks.
\newblock {\em Physical Review X}, 10(3):031044.

\bibitem{strogatz2022fifty}
Steven Strogatz, Sara Walker, Julia~M Yeomans, Corina Tarnita, Elsa Arcaute,
  Manlio De~Domenico, Oriol Artime, and Kwang-Il Goh.
\newblock (2022).
\newblock Fifty years of ‘more is different’.
\newblock {\em Nature Reviews Physics}, pages 1--3.

\bibitem{tao2016finite}
Terence Tao.
\newblock (2016).
\newblock Finite time blowup for an averaged three-dimensional navier-stokes
  equation.
\newblock {\em Journal of the American Mathematical Society}, 29(3):601--674.

\bibitem{tao2017universality}
Terence Tao.
\newblock (2017).
\newblock On the universality of the incompressible euler equation on compact
  manifolds.
\newblock {\em arXiv preprint arXiv:1707.07807}.

\bibitem{tao2019universality}
Terence Tao.
\newblock (2019).
\newblock On the universality of the incompressible euler equation on compact
  manifolds, ii. non-rigidity of euler flows.
\newblock {\em arXiv preprint arXiv:1902.06313}.

\bibitem{truesdell1960program}
Clifford Truesdell.
\newblock (1960).
\newblock A program toward rediscovering the rational mechanics of the age of
  reason.
\newblock {\em Archive for history of exact sciences}, 1(1):3--36.

\bibitem{wolpert2008physical}
David~H Wolpert.
\newblock (2008).
\newblock Physical limits of inference.
\newblock {\em Physica D: Nonlinear Phenomena}, 237(9):1257--1281.

\end{thebibliography}

\end{document}